\documentclass[12pt,preprint]{aastex} 
\shorttitle{Hot Subdwarf Companion to the Be Star 59~Cyg} 
\shortauthors{Peters et al.} 
 
\begin{document} 
 
\received{2012 November 06} 
\accepted{} 
 
\title{FUV Detection of the Suspected Subdwarf Companion \\ to the Be Star 59~Cygni}  
 
\author{Geraldine J. Peters\altaffilmark{1,2}} 
\affil{Space Sciences Center, University of Southern California, Los Angeles, CA 90089-1341, USA;  
gpeters@usc.edu} 
 
\author{Tiffany D. Pewett, Douglas R. Gies, Yamina N. Touhami} 
\affil{Center for High Angular Resolution Astronomy and  
 Department of Physics and Astronomy,\\ 
 Georgia State University, P. O. Box 4106, Atlanta, GA 30302-4106, USA; \\ 
 pewett@chara.gsu.edu, gies@chara.gsu.edu, yamina@chara.gsu.edu} 
 
\author{Erika D. Grundstrom\altaffilmark{2}}
\affil{Physics and Astronomy Department,
 Vanderbilt University, 6301 Stevenson Center, Nashville, TN 37235, USA; \\
 erika.grundstrom@vanderbilt.edu}

\altaffiltext{1}{Guest Observer with the {\it International Ultraviolet Explorer 
Satellite}.} 
 
\altaffiltext{2}{Visiting Astronomer, Kitt Peak National Observatory,
National Optical Astronomy Observatory, operated by the Association
of Universities for Research in Astronomy, Inc., under contract with
the National Science Foundation.}

\slugcomment{Version 2; submitted to ApJ} 
\paperid{89833}

%%%%%%%%%%%%%%%%%%%%%%%%%%%%%%%%%%%%%%%%%%%%%%%%%%%%%%%%%%%%%% 
 
\begin{abstract} 
We report on the detection of a hot subdwarf component in the 
Be binary system, 59~Cygni.  The spectral signature is found in 
cross-correlation functions of photospheric model spectra with 
far-ultraviolet spectra obtained by the {\it International Ultraviolet 
Explorer Satellite}, and we used radial velocities from the 
cross-correlation functions to determine a double-lined 
spectroscopic orbit.  The individual spectra of the binary components
were extracted using a Doppler tomography algorithm.  The flux 
of the system is dominated by the rapidly rotating Be star. 
However, the subdwarf contributes approximately $4\%$ of the UV flux, 
and its spectrum bears a strong resemblance to that of the hot 
sdO star BD$+75^\circ 325$.  Based upon the appearance of the 
UV spectrum and the orbital elements, we present estimates for 
the stellar masses, radii, and temperatures.   The presence of
the hot companion causes excess emission from the outer part of
the Be disk facing the companion.  We present a set of red 
spectra that show the orbital phase variations of the \ion{He}{1}
$\lambda 6678$ emission formed in the heated region of the disk, 
which probably occurs near the disk outer boundary. 
59~Cygni, FY~Canis Majoris, and $\phi$~Persei comprise 
the known set of Be binaries with detected hot
evolved companions, which are the stripped down remains of 
mass transfer.  Their properties demonstrate that some fraction of 
Be stars were spun up through angular momentum transfer by 
Roche lobe overflow. 
\end{abstract} 
 
 
%%%%%%%%%%%%%%%%%%%%%%%%%%%%%%%%%%%%%%%%%%%%%%%%%%%%%%%%%%%%%%% 
 
\setcounter{footnote}{0} 
 
\section{Introduction}                              % Section 1 
 
The origin of the rapid rotation that characterizes the 
emission line Be stars is still unknown \citep{por03}.
One important clue is that Be stars are frequently the 
primary components of massive X-ray binaries with 
neutron star companions \citep{rei11}, and thus, many are 
members of binaries that must have interacted in the past
\citep{pol91}.  Past mass transfer from the neutron star 
progenitor to the mass gainer probably resulted in a 
spin up of the gainer to near critical rotation \citep{pac81}, 
and the current mass loss into the circumstellar disk of the Be star
represents processes that shed some of the star's
extreme angular momentum.  In many cases, the donor 
star may lose enough mass to drop below the Chandrasekhar 
limit, and the result may be a hot helium star (spectral type sdO) 
or (eventually) a white dwarf remnant \citep{gie00}.  
Such faint companions are very difficult to detect, but
because they are hot, searches in the ultraviolet part
of the spectrum are favored, and we now have found the UV 
spectral signatures of hot helium cores in two Be binaries, 
$\phi$~Per \citep{gie98} and FY~CMa \citep{pet08}.  

Here we turn our attention to the remarkable collection 
of UV spectra in the archive of the {\it International 
Ultraviolet Explorer (IUE)} of the Be binary 
59~Cygni (HD~200120, HR~8047, V832~Cyg).  The star is the 
brightest component of a multiple system with a nearby companion Ab 
detected through speckle interferometry \citep{mas09}.  An even 
closer binary companion of the central star Aa was suspected 
for a long time, and the first orbital determinations were made 
by \citet{tar87} and \citet{riv00} who correctly 
identified the 28~d orbital period of the binary.
\citet{har02} presented a joint photometric and 
spectroscopic analysis, and they determined orbital 
elements from radial velocity measurements of the 
H$\alpha$ emission line wings (formed in the disk close to the Be star). 
\citet{mai03} and \citet{mai05} presented additional radial velocity 
measurements for the Be star, and they showed that anti-phase 
Doppler shifts are present in the \ion{He}{2} $\lambda 4686$ absorption 
line, a feature only found in stars much hotter than the Be star
primary (classified as B1.5~Vnne by \citealt{les68}). 
Maintz and collaborators argued that the \ion{He}{2} line forms in 
the atmosphere of a hot, faint companion and that this companion must 
be the stripped-down remains of the former mass donor star. 
They also documented how the hot companion illuminates the 
nearby rim of the Be star's disk and causes orbital-related  
emission line variations similar to those observed in the 
spectra of $\phi$~Per and FY~CMa.  

The spectral lines of a hot companion should be present in the 
ultraviolet spectrum, but they may be difficult to discern 
in individual spectra because of the complex line blending 
in the UV spectra of hot stars and the relative faintness of
the companion compared to the Be star.  The general appearance and long-term
variations of the strong wind lines in the UV spectrum of 59~Cyg are 
described in detail by \citet{doa89}.  Here we apply the   
cross-correlation method to detect the companion spectrum and to measure
its radial velocity variations \citep{gie98,pet08}.  The measurements and 
the derivation of orbital elements are described in Section 2, and 
we use the radial velocity curves and a Doppler tomography algorithm 
to reconstruct the spectra of both components in Section 3.
We describe in Section 4 how the flux of the subdwarf companion 
heats the nearby region of the Be star disk and creates 
orbital phase related emission line variations.  
We summarize these new results on 59~Cyg and their relation to 
other Be + sdO binaries in Section 5. 
 
%%%%%%%%%%%%%%%%%%%%%%%%%%%%%%%%%%%%%%%%%%%%%%%%%%%%%%%%%%%%%%% 
 
\section{Radial Velocities and Orbital Elements}    % Section 2 

The archival spectra of 59~Cyg from observations made with the 
{\it International Ultraviolet Explorer} provide a valuable record
of the orbital Doppler shifts associated with the binary motion. 
Here we describe the observational properties of the spectra, 
and then we outline our use of the cross-correlation method to 
measure radial velocities of first the hot secondary star and 
then the relatively cooler Be star.  Because the radial velocity 
semiamplitude is much larger for the hotter star, we begin the 
fit of the orbital elements with the hot star velocities and 
then make a restricted fit of the associated elements for 
the Be star.  We conclude this section with a comparison to 
earlier published work. 

We obtained the UV spectra of 59~Cyg from the NASA Mikulski Archive 
for Space Telescopes\footnote{http://archive.stsci.edu/iue/search.php}.
The selection focused on observations made with the {\it IUE} HIRES, 
Short Wavelength Prime (SWP) camera, which recorded the far-UV spectrum
with a resolving power of $R=10000$.  There are a total of 194 such 
observations that were made between 1978 and 1994.  We further limited
this sample to the 157 spectra with the best S/N ratio (usually $>10$ 
in the better exposed parts of the spectrum).   The echelle orders
were extracted and combined using pipeline IUERDAF software, and 
then each spectrum was normalized to unity in relatively line-free 
regions and transformed to a standard, heliocentric wavelength grid 
in $\log \lambda$ that spans from 1200 to 1900 \AA . 
The wavelength calibration of each spectrum was checked for possible offsets 
due to the star's position in the large aperture by comparing the 
positions of the strong interstellar lines with those in the 
global average spectrum.  For each spectrum, we made cross-correlation 
measurements of the pixel shifts between the strong interstellar lines
in a given spectrum with those in the average of all the spectra, 
and then we applied the median of these measurements to shift the
observed interstellar spectrum into alignment with the average 
interstellar spectrum.  Then the interstellar lines were removed from 
each spectrum by replacement with a linear interpolation to the 
spectrum at the boundaries of each interstellar feature. 

Our primary goal is to detect the spectrum of the hot companion 
in order to measure the star's radial velocity.  It is very difficult
to measure the position of any particular line of the companion's 
spectrum because of the limited S/N of the observations and 
the small fraction of the companion's flux (see below).  
However, we can measure the signal of the ensemble of spectral 
lines of the companion by calculating the cross-correlation function 
(ccf) of the spectrum with that of a model for the hot star. 
This approach is successful because the companion is so much hotter
than the Be star that almost all of its spectral features are comprised
of higher ionization species, and consequently the ccf signal is 
only sensitive to the lines of the hot companion and not to those
of the cooler Be star.  We constructed a hot star spectrum template
from the spectral models of \citet{lan03} for $T_{\rm eff} = 55000$~K, 
$\log g = 4.4$, $V\sin i = 0$ km~s$^{-1}$, and solar abundances. 
The model was transformed to the observed wavelength scale by 
flux integration into the same $\log \lambda$ wavelength system, and
then the model was convolved with the instrumental broadening function.
The ccf was calculated over the entire spectrum after first setting 
to unity the low wavelength end (1200 -- 1226 \AA\ to avoid Ly$\alpha$)
and those regions in the immediate vicinity of strong wind lines or 
interstellar lines.  Consequently, the results are largely free from 
the influence of non-photospheric spectral features. 

We show the results of the ccf calculations in Figure~1 as a
function of radial velocity and orbital phase.  The upper 
panel illustrates the difference between each ccf and the mean of 
all the ccfs in order to remove the broad and constant background 
ccf structure caused by low frequency correlation with the Be star spectrum. 
It is hard to find the peak signal from the spectrum of the hot
companion in individual ccfs because the peaks have an amplitude 
not much larger than the noise.  However, the signal is much more 
evident in the lower panel where the ccf differences are plotted 
as a gray scale intensity as a function of radial velocity and
orbital phase.  The bright, ``S''-shaped curve reveals the orbital 
motion of the hot companion, despite the weakness of the signal, 
thanks to the large number of observations available. 

\begin{figure}
\begin{center} 
{\includegraphics[height=16cm]{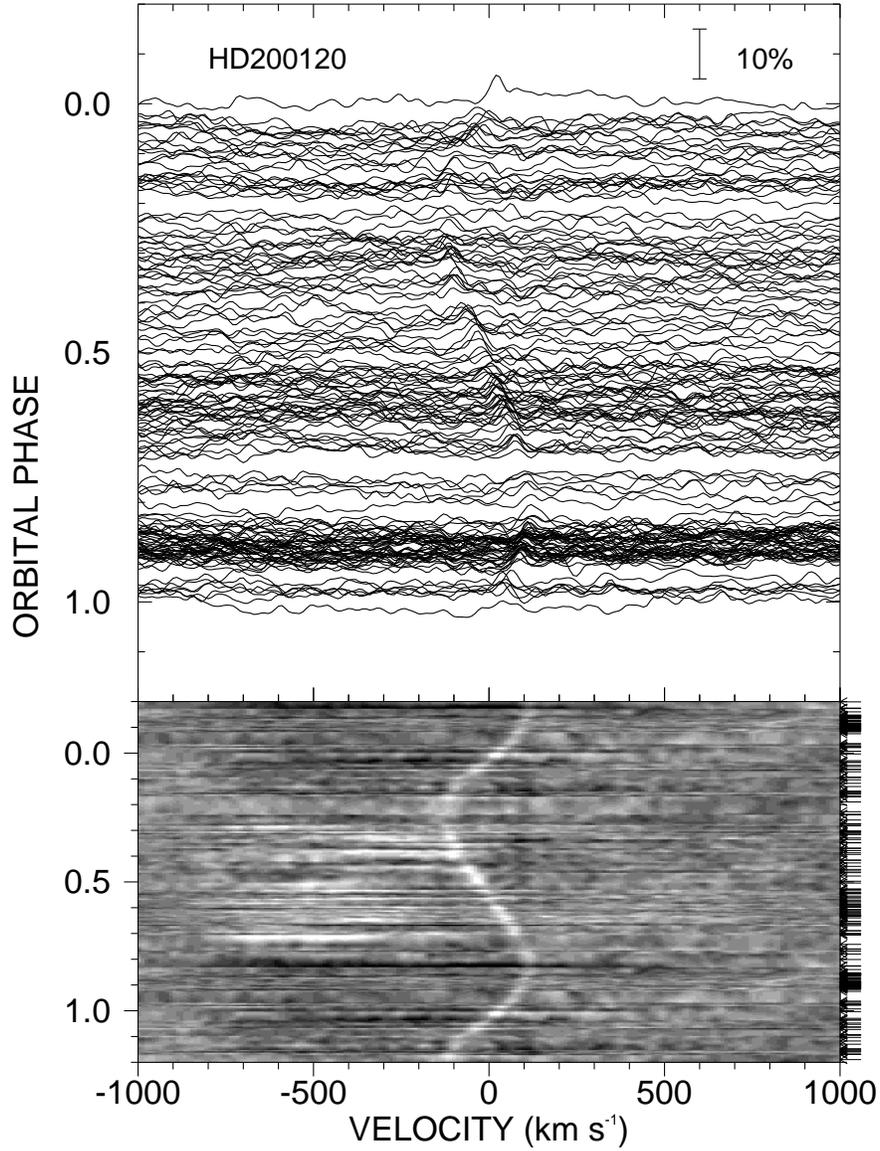}}
\end{center} 
\caption{The orbital phase variations of the cross-correlation functions 
with the hot star template spectrum after subtraction of the mean ccf.  
The ccf differences are shown as linear plots ({\it top panel}) 
and as a gray-scale image ({\it lower panel}).  
The intensity in the gray-scale image is scaled to $\pm 4\%$ and  
is calculated by a linear interpolation between the closest observed phases 
(shown by arrows along the right axis).  The weak signal from the hot subdwarf 
spectrum appears as the ``S'' feature in the lower panel.  
\label{fig1}} 
\end{figure}

\placefigure{fig1}     % Figure 1 - Subdwarf ccf plot  
 
We estimated the radial velocities of the hot companion 
by fitting a parabola to the central $\pm 20$ km~s$^{-1}$ around 
the peak of each ccf.  This was not always successful 
because the peak was too weak in many cases, and in the 
end we restricted the results to 132 spectra where a 
reliable measurement was possible.   The results are 
summarized in Table~1 (given in full in the electronic version)
that lists the heliocentric Julian date of mid-observation, 
the corresponding UT date, the {\it IUE} SWP number assigned 
to the spectrum, and orbital phase (see below).  
Columns 5 -- 7 and 8 -- 10 then give the measured 
radial velocity ($V$), uncertainty ($\sigma$), and 
observed minus calculated residual ($O-C$) for both the
primary Be star (see below) and hot companion, respectively. 
Note that in most cases that the formal uncertainty $\sigma$
(based in the formulation of \citealt{zuc03}) is smaller than  
the absolute value of the residual $O-C$ because the 
former does not account for uncertainties associated with 
the wavelength calibration of each spectrum. 

\placetable{tab1}      % Table 1 - IUE ccf velocities 
  
We then fit the radial velocities to obtain orbital elements 
using the non-linear, least-squares solver of \citet{mor74}.
The solution was obtained by assigning equal weights to 
each measurement, which is appropriate for the uniform S/N 
properties of the final set of measurements.  The resulting 
elements and their uncertainties are given in column 4 of 
Table 2, where the subscript 2 is used to identify parameters
for the hot companion.  These include the orbital period $P$, 
the epoch of periastron $T$, the epoch of Be star superior 
conjunction $T_{SC}$ (equal to the epoch of inferior 
conjunction for the hot companion), the velocity semiamplitude $K$, 
the systemic velocity $\gamma$, the eccentricity $e$, 
the longitude of periastron $\omega$, and the root mean square of 
the residuals from the fit.  We adopt the standard orbital phase 
convention in which orbital phase zero corresponds to the time of periastron. 
The orbital velocity curve and measured radial velocities are plotted 
together in Figure 2.   Note that the fitting scheme uses a version of 
the Levenberg-Marquardt method that estimates the parameter errors 
according to the covariance matrix and the final $\chi^2$ of the fit. 

\placetable{tab2}      % Table 2 - Orbital elements 
 
\begin{figure} 
\begin{center} 
{\includegraphics[angle=90,height=12cm]{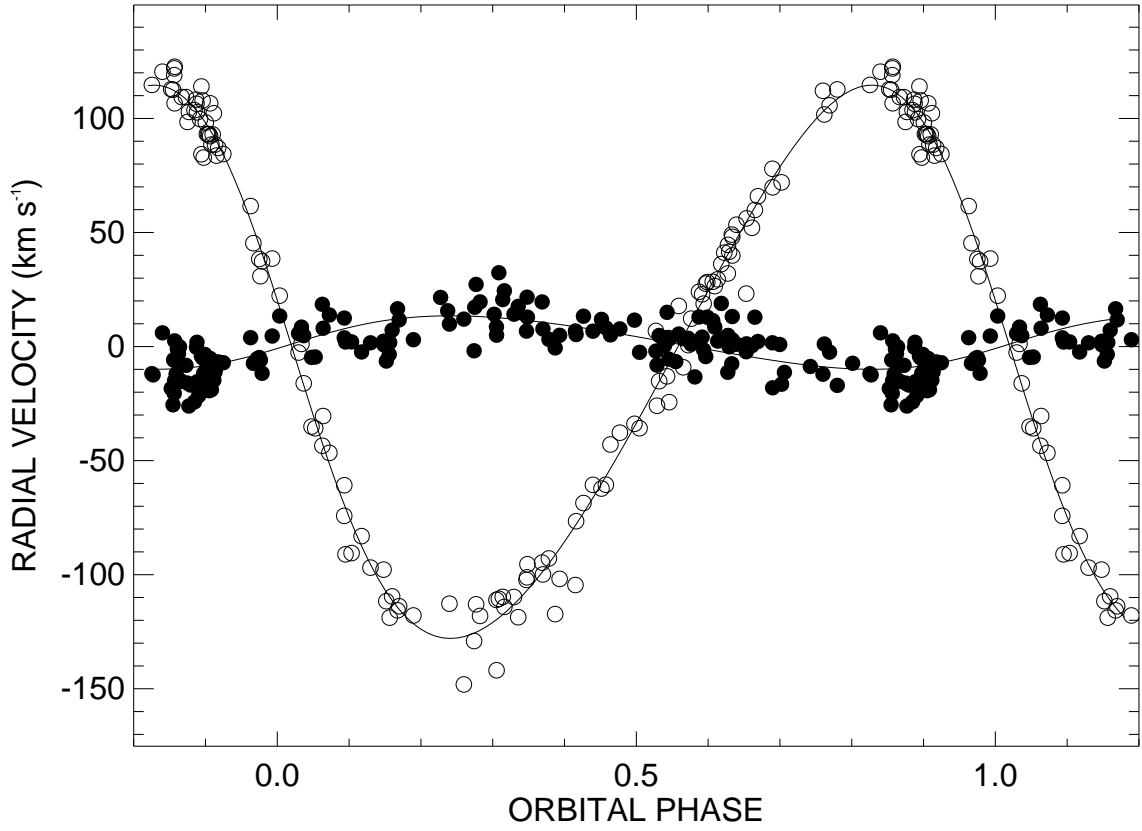}} 
\end{center} 
\caption{Radial velocity curves for the Be star and its companion. 
Orbital phase 0.0 corresponds to periastron.  
The solid circles represent the ccf velocities for the Be star while 
the open circles show the same for the hot subdwarf.  
\label{fig2}} 
\end{figure}

\placefigure{fig2}     % Figure 2 - Velocity curves 
 
We also used the cross-correlation method to measure radial 
velocities for the Be star primary, but in this case we 
used the average of all the spectra as the spectrum template. 
This is a reasonable choice because the flux of the Be star
dominates the far UV spectrum.  The resulting ccfs show a 
very broad peak that reflects the large rotational broadening 
of the primary star's spectrum plus a weak and narrow central
peak.  We suspect that the latter component may result from
residual interstellar lines, edges corresponding to cuts of
the stronger interstellar features, and/or lines of 
the relatively stationary third component, Ab \citep{mas09}. 
Thus, instead of fitting a parabola to the central part, 
we measured the ccf positions using the bisector of the wings that 
was calculated using the method of \citet{sha86}.   
The transformation from relative to absolute radial velocity requires 
an estimate for the velocity of the average spectrum.  
We found that the velocity offset between the average spectrum 
and a model (for $T_{\rm eff} = 21750$~K, $\log g = 3.8$, 
$V\sin i = 379$ km~s$^{-1}$; \citealt{fre05}) was $7.8 \pm 7.4$
km~s$^{-1}$ (by the wing bisector method).  However, this ccf
with the model was visibly asymmetrical, and the resulting bisector
velocity depended on the part of the wing sampled.  Consequently, 
the transformation from relative to absolute velocity remains 
problematical, so no correction was applied to the 
velocities of the primary.  The resulting relative velocities, 
uncertainties, and residuals are listed in columns 5 through 7
of Table~1 for all 157 spectra selected. 

A first fit of the primary velocities with only the period fixed
resulted in estimates of $e$, $\omega_1 = \omega_2 +180^\circ$, 
and $T$ consistent within uncertainties with the respective
values for the secondary but with much larger values of uncertainty. 
Thus, we simply fixed $P$, $T$, $e$, and $\omega$ to those 
values derived from the larger amplitude radial velocity curve
of the secondary, and then solved for $K_1$ and $\gamma_1$ only. 
The derived parameters are listed in column 4 of Table 2, 
and the radial velocity curve and measurements are over-plotted
with those for the secondary in Figure 2.  With estimates in hand
for the parameters of both components, we can then 
determine the mass ratio, $M \sin^3 i$ products, and the combined
projected semimajor axis $a\sin i$, which are all listed 
in Table 2.

Our results for the radial velocity curve of the hot companion
are in excellent agreement with those from \citet{mai03} and 
\citet{mai05}, and this confirms their conclusion that the 
Doppler shifts of the \ion{He}{2} $\lambda 4686$ line trace 
the motion of the hot star.   On the other hand, our estimate for 
the semiamplitude of the Be star is much smaller than that found 
by Maintz and collaborators, but agrees with the determination 
from \citet{har02} based on the Doppler shifts of the H$\alpha$
emission line wings.  We think that the difference arises because
of residual emission in the \ion{He}{1} lines measured by 
\citet{mai05}.  They point out that the emission component 
formed in the outer disk facing the hot companion 
(which is seen dramatically in \ion{He}{1} $\lambda 6678$; Section 4) 
will probably partially fill in the absorption profiles of the other 
\ion{He}{1} lines in such a way as to increase the apparent 
Doppler shifts.  The UV lines, on the other hand, correspond to larger 
energy transitions than those of \ion{He}{1}, and the temperatures 
in the heated portion of the disk are probably insufficient 
to create any significant emission line flux in the FUV 
(and we see no evidence of such emission in the {\it IUE} spectra). 
The fact that the velocity semiamplitude $K_1$ we derive from the UV 
photospheric spectrum is so similar to that from the H$\alpha$ emission wings
(formed in the inner part of the Be star's disk) is consistent 
with the assumption that the Doppler shifts we measure correspond
to those of the Be star itself.  Estimates of the other orbital elements 
(with the possible exception of $\omega$) are consistent with those 
from \citet{har02}, \citet{mai03}, and \citet{mai05}. 

%%%%%%%%%%%%%%%%%%%%%%%%%%%%%%%%%%%%%%%%%%%%%%%%%%%%%%%%%%%%%%% 
 
\section{Tomographic Reconstruction of the UV Spectra}  % Section 3 
 
The nature of the faint companion star can be analyzed through 
an examination of its UV spectrum.  However, because the companion is 
so faint compared to the Be primary, its spectrum cannot be easily
extracted from any individual spectrum.  Instead, we used a 
Doppler tomography algorithm \citep{bag94} to extract the 
component spectra, and because this approach uses all the 
spectra, the reconstructed spectra have 
a S/N ratio improved by $\approx \sqrt{n}$, where $n$
is the number of spectra.  We used the orbital solutions to 
determine the radial velocities of the primary and secondary for 
each spectrum, and we adopted a small flux fraction to begin (that
was adjusted later to match spectral depths in the reconstructed 
spectra; see below).  We adopted the same model spectra used for 
ccf analysis as starting values and we ignored the flux contribution 
from the faint Ab speckle companion.  We ran the Doppler tomography
for 50 iterations with a gain of 0.9, and tests showed that the resulting
spectra were insensitive to all these assumptions.  The reconstructed 
spectrum of the hot secondary shows considerable wander in the continuum 
on scales of tens of \AA ngstroms, so we reset the continuum by forming 
Gaussian smoothed (FWHM = 610 km~s$^{-1}$) versions of both the 
reconstructed and model spectra and then by dividing the reconstructed
spectrum by the ratio of the smoothed spectra.  This method removes
all long wavelength span features but retains the narrow features that
characterize the spectrum of the secondary.  The rectified version 
of the reconstructed secondary spectrum is illustrated in Figures 3 to 7
that show spectral subregions containing lines of the principle ions 
that characterize the spectrum.   

\begin{figure} 
{\includegraphics[angle=90,height=12cm]{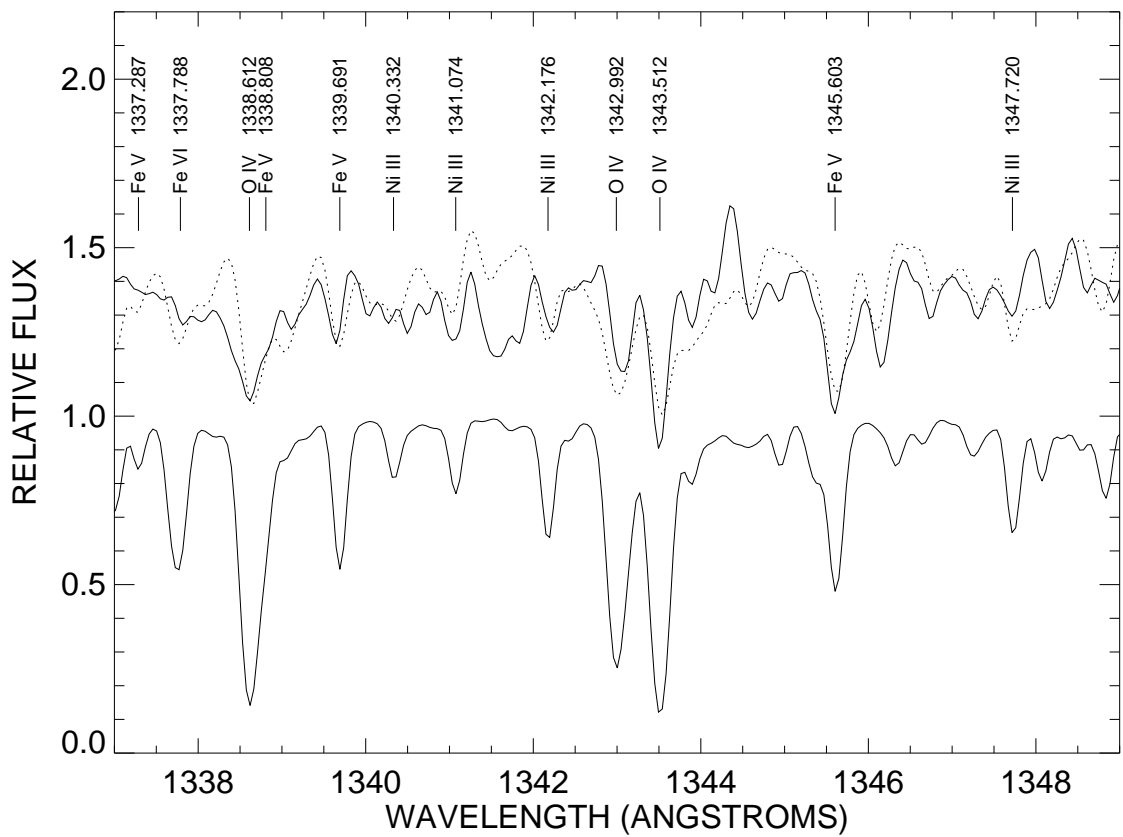}} 
\caption{A comparison of the \ion{O}{4} lines in the reconstructed UV spectrum of the 
secondary star ({\it upper solid line}) with those in the average spectrum of BD+75$^\circ$325 
({\it dotted line}). Both are offset vertically by +0.5 to separate them 
from the model spectrum below (shown with no offset).   All the spectra were 
smoothed by convolution with a Gaussian function with FWHM = 30 km~s$^{-1}$ (3 pixels). 
Line identifications for some of the stronger lines are given above the spectra. 
\label{fig3}} 
\end{figure}

\placefigure{fig3}     % Figure 3 - Sec. reconstruction 
 
\begin{figure} 
{\includegraphics[angle=90,height=12cm]{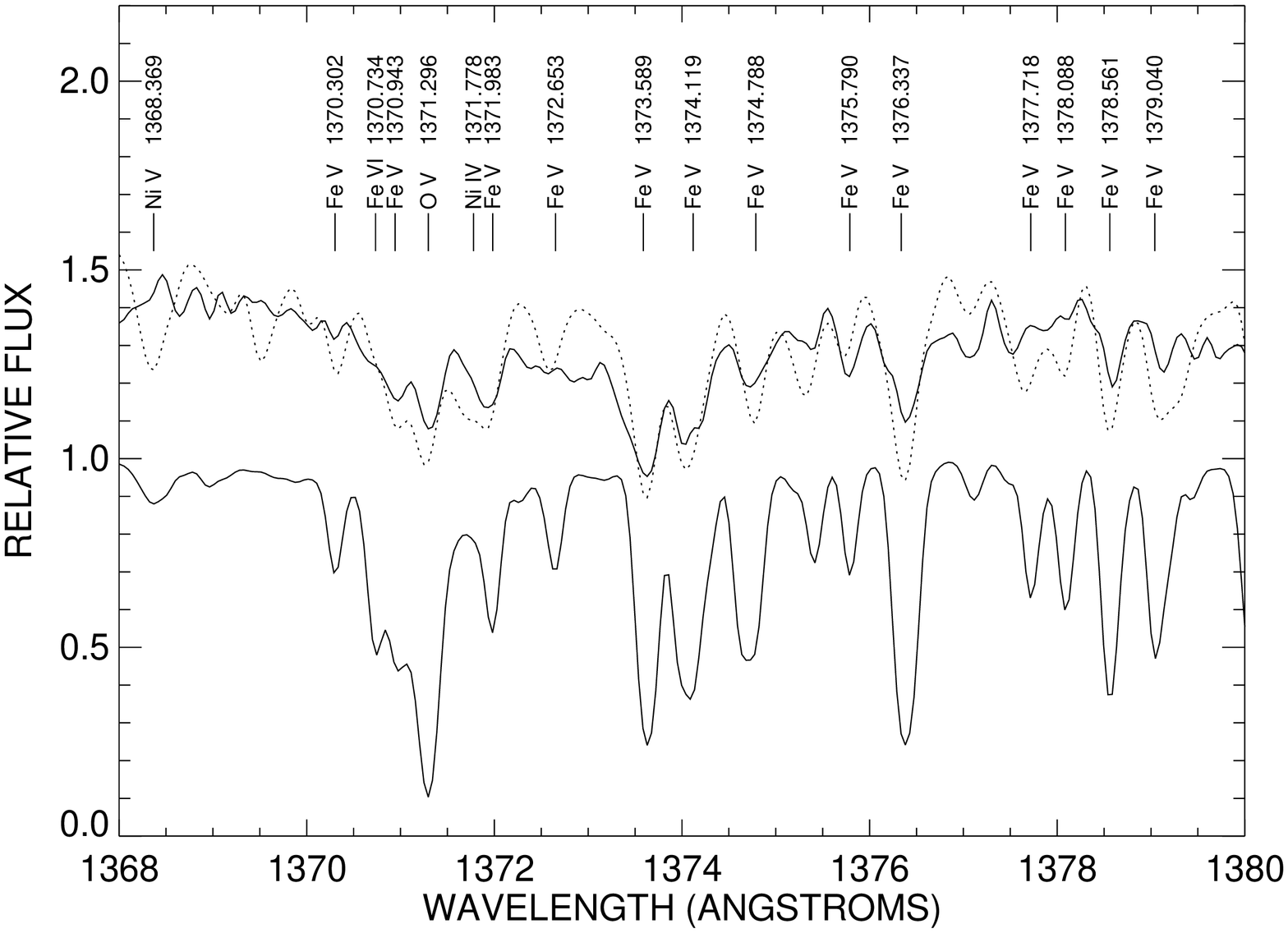}} 
\caption{A comparison of the \ion{O}{5} and \ion{Fe}{5} lines in the reconstructed UV spectrum of the 
secondary star ({\it upper solid line}) with those in the average spectrum of BD+75$^\circ$325 
({\it dotted line}) in the same format as Fig.~3. 
\label{fig4}} 
\end{figure}

\placefigure{fig4}     % Figure 4 - Sec. reconstruction 
 
\begin{figure} 
{\includegraphics[angle=90,height=12cm]{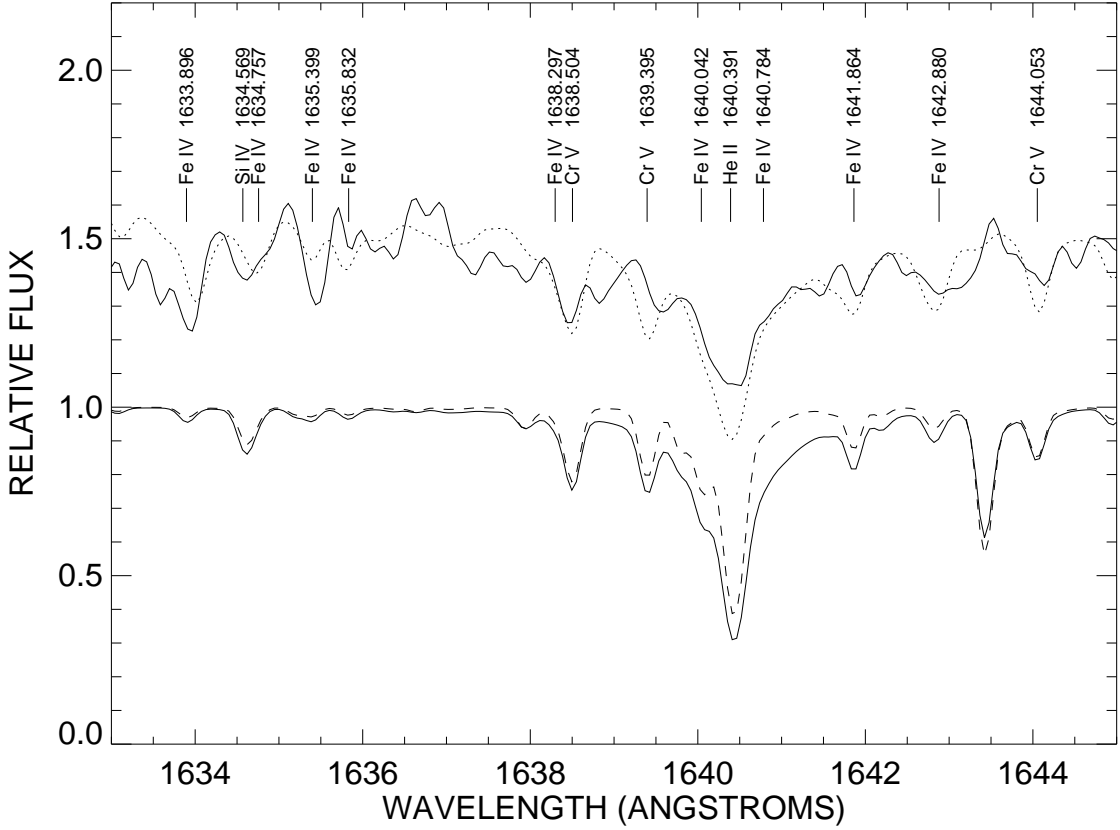}} 
\caption{A comparison of the \ion{He}{2} line in the reconstructed UV spectrum of the 
secondary star ({\it upper solid line}) with those in the average spectrum of BD+75$^\circ$325 
({\it dotted line}) in the same format as Fig.~3.  The two lines in the lower 
part show how the \ion{He}{2} line wings become narrower at $\log g = 4.00$ ({\it dashed line}) 
compared to the nominal case with $\log g = 4.75$ ({\it solid line}).  
\label{fig5}} 
\end{figure}

\placefigure{fig5}     % Figure 5 - Sec. reconstruction 
 
\begin{figure} 
{\includegraphics[angle=90,height=12cm]{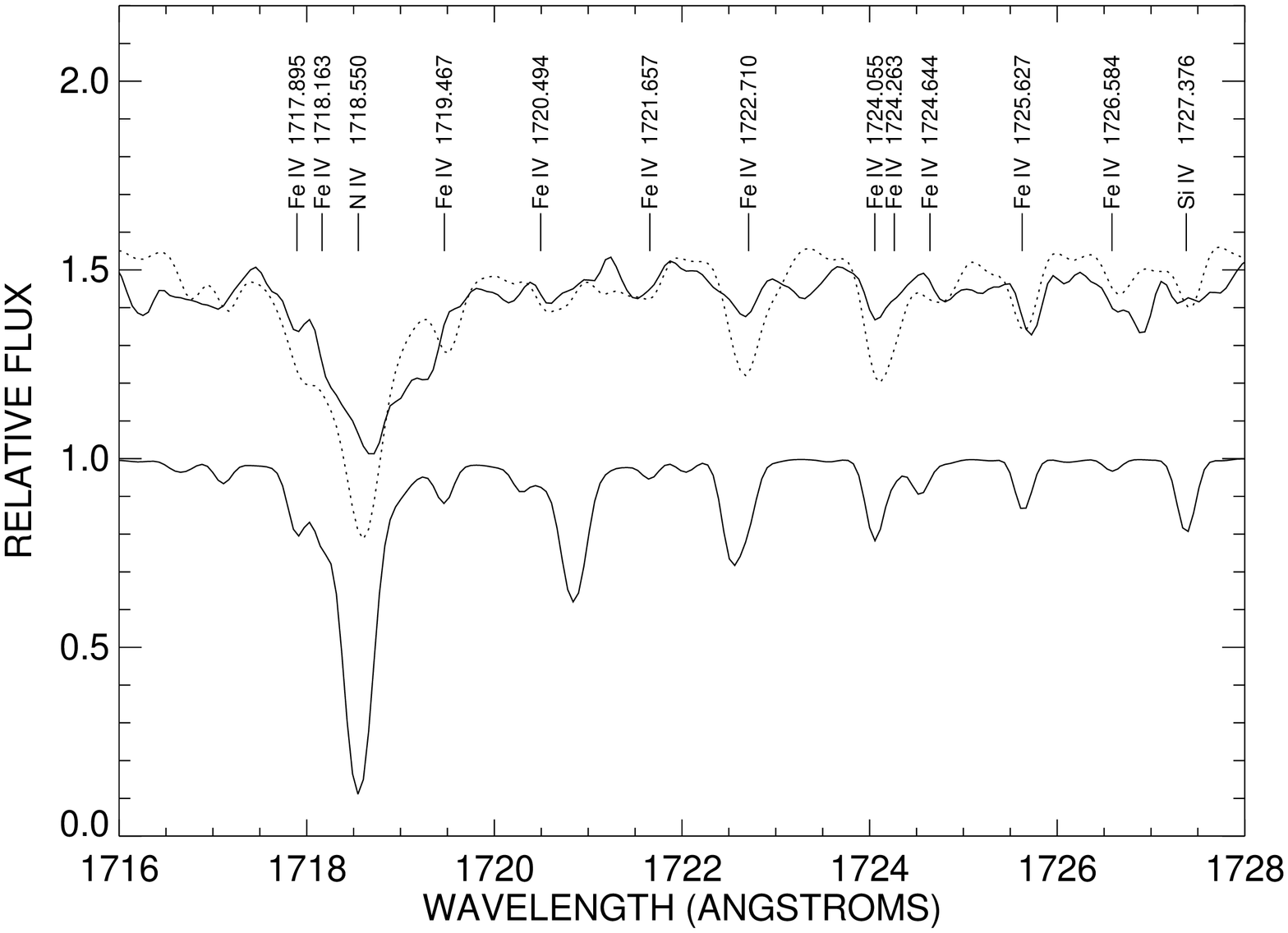}} 
\caption{A comparison of the \ion{N}{4} and \ion{Fe}{4} lines in the reconstructed UV spectrum of the 
secondary star ({\it upper solid line}) with those in the average spectrum of BD+75$^\circ$325 
({\it dotted line}) in the same format as Fig.~3. 
\label{fig6}} 
\end{figure}

\placefigure{fig6}     % Figure 6 - Sec. reconstruction 
 
\begin{figure} 
{\includegraphics[angle=90,height=12cm]{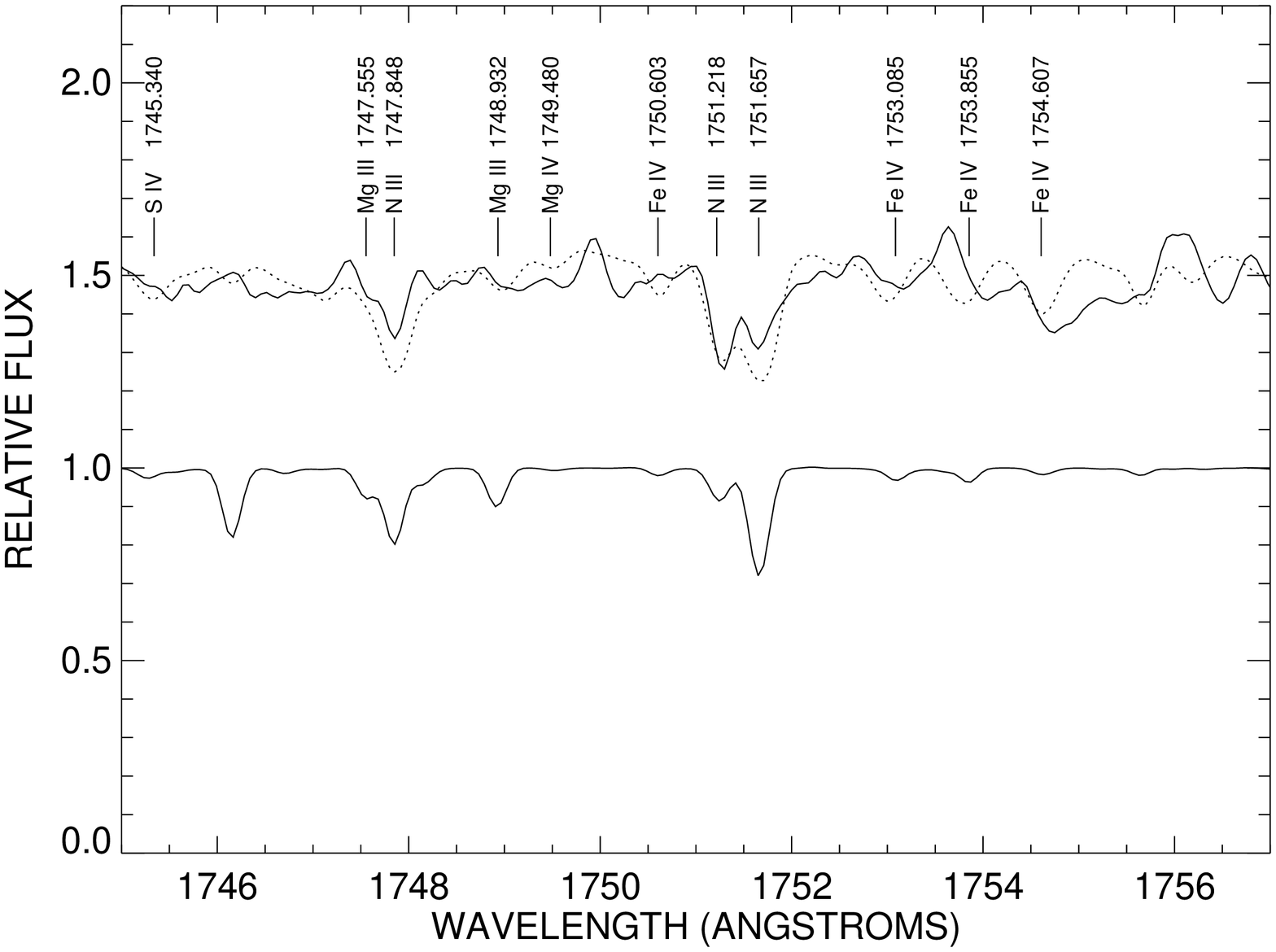}} 
\caption{A comparison of the \ion{N}{3} lines in the reconstructed UV spectrum of the 
secondary star ({\it upper solid line}) with those in the average spectrum of BD+75$^\circ$325 
({\it dotted line}) in the same format as Fig.~3. 
\label{fig7}} 
\end{figure}

\placefigure{fig7}     % Figure 7 - Sec. reconstruction 
 
We compared the reconstructed spectra to TLUSTY/SYNSPEC models 
from \citet{lan07} for the Be star primary and from \citet{lan03}
for the hot secondary.  The model for the primary based on 
the parameters from \citet{fre05} (see Table 3) makes an 
excellent match of the reconstructed primary spectrum. 
To find the best match of the reconstructed secondary spectrum, 
we created a grid of model UV spectra and cross-correlated 
each one with the reconstructed secondary spectrum.  The largest 
amplitude ccf peak was found at an interpolated effective temperature 
of 47.3 kK for models with $\log g = 4.00$ and at 52.1 kK for models
with $\log g = 4.75$ (the highest value available in the published
grid of \citealt{lan03}).  This $(T_{\rm eff}, \log g)$ relation  
presumably makes the best overall match of the \ion{Fe}{4} to 
\ion{Fe}{5} line depth ratios (and of those for other prominent ions). 
The ccf peak is slightly higher at the highest gravity, which is consistent
with the appearance of the pressure broadened wings of 
\ion{He}{2} $\lambda 1640$ (Fig.\ 5).  We caution, however, that 
the models from \citet{lan03} assume solar abundances and a microturbulent
velocity of 10 km~s$^{-1}$, and changes about these assumptions 
will influence the derived best fit temperature.  The model spectrum for  
$T_{\rm eff} = 52.1$ kK and $\log g = 4.75$ is shown offset below the 
reconstructed secondary spectrum in Figures 3 to 7. 

\placetable{tab3}      % Table 3 - Physical properties 

Another comparison can be made to the observed UV spectrum of the 
single, sdO star, BD$+75^\circ 325$.  The spectrum of this star 
was analyzed in detail by \citet{lan97} who used the TLUSTY/SYNSPEC 
models to derive a helium abundance (He:H = 1 by number), $T_{\rm eff} = 52$ kK, 
$\log g = 5.5$, and a microturbulent velocity of 10 km~s$^{-1}$. 
A comparison with BD$+75^\circ 325$ is particularly relevant because the hot 
companion in 59~Cyg is probably He-enriched due to past stripping by mass transfer.
Fortunately, there are 92 UV spectra of BD$+75^\circ 325$ in the {\it IUE} 
archive that were made with the same SWP, high dispersion camera.  
We obtained these and processed them in the same way as done for the 
59~Cyg spectra.  Each spectrum was cross-correlated with a model 
TLUSTY/SYNSPEC spectrum to obtain the radial velocity ($<V_r> = -59.4 \pm 0.3$ 
km~s$^{-1}$), and then each spectrum was shifted to the lab frame before 
co-addition.  Finally, the rectification of the mean spectrum was adjusted in 
the same way as done for the reconstructed secondary spectrum in order to 
inter-compare the spectra.  The final UV spectrum of BD$+75^\circ 325$ 
is over-plotted in Figures 3 to 7, and, in general, it makes an excellent 
match to that of the hot secondary.  

The photospheric lines of both the hot secondary and BD$+75^\circ 325$
appear very sharp, and comparisons with model spectra indicate that 
any rotational line broadening is unresolved ($V\sin i < 40$ km~s$^{-1}$). 
Consequently, if we assume that the spectrum of BD$+75^\circ 325$ 
is an effective match of that of the secondary, then we can find the 
monochromatic (mean UV) flux ratio $f_2/f_1$ of the secondary to primary flux 
by adjusting the flux ratio used in the tomographic reconstruction so that the 
resulting line depths match those in the spectrum of BD$+75^\circ 325$.
We derived this flux ratio by comparing the line depths in 
smoothed versions of the spectra in nine regions of 
$\approx 50$ \AA\ width that contained significantly deep lines. 
The best fit was obtained with a flux ratio $f_2/f_{\rm total} = 0.038 \pm 0.005$, 
and this value was used in the tomographic reconstruction shown in 
Figures 3 to 7.  The formal error does not account for changes  
in line depth due to differences between secondary and BD$+75^\circ 325$
related to $T_{\rm eff}$, $\log g$, abundances, and microturbulence.
Our impression from the good match made with the spectrum of BD$+75^\circ 325$
is that such differences are probably minor.  Our results are summarized 
in Table~3 that lists $T_{\rm eff}$, $\log g$, and $V\sin i$ estimates 
for the hot subdwarf (column 3) along with those for the Be star (column 2)
from \citet{fre05}.

%%%%%%%%%%%%%%%%%%%%%%%%%%%%%%%%%%%%%%%%%%%%%%%%%%%%%%%%%%%%%%% 
 
\section{Heating of the Disk by the Companion}      % Section 4 

One other means of detecting a hot companion is through 
observations of line emission from a heated region of the 
Be star disk that faces the companion \citep{hum01,hum03}.  
The localized heating effects are especially striking 
in the weak emission line of \ion{He}{1} $\lambda 6678$ 
that usually displays a radial velocity curve similar to that 
of the companion but with a significantly larger semiamplitude. 
The \ion{He}{1} $\lambda 6678$ emission variations are 
evident in the spectra of the known Be + sdO binaries,
59~Cyg \citep{mai05}, $\phi$~Per \citep{gie93,ste00}, and 
FY~CMa \citep{pet08}, and their presence is the basis of the 
proposed sdO companion of the Be star $o$~Pup \citep{kou12,riv12}.

Here we revisit the \ion{He}{1} $\lambda 6678$ emission line 
variations in the spectrum of 59~Cyg using a series of spectra 
we obtained with the Kitt Peak National Observatory Coud\'{e}
Feed Telescope and additional spectra from the Database of 
Be Star Spectra\footnote{http://basebe.obspm.fr/basebe/} (BeSS) \citep{nei11}.
Table~4 lists the origin, time span, spectral resolving power, 
number of observations, and observer name for the 59 spectra collected. 
All these spectra have a S/N ratio $>200$, a resolving power 
greater than $\approx 10000$, and record most of the 
\ion{He}{1} $\lambda 6678$ feature (although in some of the 
BeSS spectra the blue or red wings are located at the boundary of 
the recorded spectrum).  All the spectra were rectified to a 
unit continuum and were transformed to a uniform wavelength 
grid in the heliocentric frame.  The spectra are presented in Figure~8 
as a function of heliocentric velocity and orbital phase (from 
periastron at phase 0).  The upper portion shows the individual 
spectral profiles with the continua aligned with the orbital phase,
and the lower portion shows the spectral fluxes as a gray-scale image
made through linear interpolation in orbital phase.  The radial 
velocity excursions of the emission feature resemble those 
of the hot component (Fig.~1) but the emission has a larger semiamplitude
($K_{\rm em} = 180 \pm 10$ km~s$^{-1}$).  The overall appearance of 
the gray-scale figure is very similar to that presented by \citet{mai05}
(see their Fig.~2, which shows orbital phase increasing upwards), and 
this demonstrates that the emission line variations have continued through 
to the present time. 

\placetable{tab4}      % Table 4 - He I 6678 observations 

The \ion{He}{1} $\lambda 6678$ emission probably originates in the
outer parts of the Be star disk that are directly illuminated by
the flux of the hot subdwarf companion.  The model presented by 
\citet{pet08} for FY~CMa (see their Fig.~8) can also be applied 
here to the case of 59~Cyg.  In the reference frame of the Be star, 
the disk gas near the outer edge has a Keplerian velocity that 
exceeds that of the companion because of its location closer to 
the Be star.  If we ignore third body effects, then we can 
estimate the approximate disk radius of line formation according to
\begin{equation} 
{{R_{\rm em}}\over{a}} = {{M_1}\over{M_1 + M_2}} \left({{K_1 + K_2} \over {K_1 + K_{\rm em}}}\right)^2 \\
= {{1}\over{1 + K_1 / K_2}} \left({{K_1 + K_2} \over {K_1 + K_{\rm em}}}\right)^2
\end{equation}
where $K_1$, $K_2$, and $K_{\rm em}$ are the semiamplitudes of the 
Be star, the subdwarf, and the \ion{He}{1} emission, respectively. 
From the estimates given above, we obtain $R_{\rm em}/a = 0.44$, 
i.e., the emission region forms about midway between the Be star
and subdwarf, presumably at a location near the outer boundary of 
the disk. 

%%%%%%%%%%%%%%%%%%%%%%%%%%%%%%%%%%%%%%%%%%%%%%%%%%%%%%%%%%%%%%% 
 
\section{Discussion}                                % Section 5 

The physical properties of the component stars are summarized 
in Table 3.   The mass estimates are derived from the 
orbital results and an orbital inclination range of $60^\circ$
to $80^\circ$ advocated by \cite{mai05} based upon the 
large projected rotational velocity and the past history of 
spectroscopic shell events.  We can estimate the radii from the 
flux ratio, the spectral energy distribution, and the distance. 
The observed flux ratio is approximately related to radius ratio by
\begin{equation}
{f_2 \over f_1} = {F_2 \over F_1} \left({R_2 \over R_1}\right)^2
\end{equation}
where $F_2 / F_1$ is the monochromatic flux ratio per unit area. 
Using the adopted temperatures in Table 3, we estimate that 
$F_2 / F_1 = 10.6$ at 1500 \AA\ based upon the TLUSTY/SYNSPEC models. 
Then the radius ratio is $R_2 / R_1 = 0.062$ after correction 
for the small flux contribution of the Ab component (assuming
$\triangle m = -2.5 \log f_{Ab}/(f_1+f_2) = 2.84$; \citealt{mas09}). 
\citet{tou12} estimated the angular diameter of primary Be star,
$\theta = 0.149 \pm 0.004$ mas, by fitting the observed FUV spectrum 
to a composite model spectrum of Aa + Ab.  This estimate should be 
considered as a value intermediate between the polar and equatorial diameters for
the rapidly rotating Be star.  Then to obtain the physical radius, 
we need to multiply the angular size times the distance. 
The original {\it Hipparcos} distance was $345 \pm 76$ pc \citep{per97},
which was revised to $435 \pm 79$ pc by \citet{van07}. 
The radii given in Table~3 correspond to a distance range of 345 to 435 pc. 
 
Hot companions of Be stars have now been detected in three systems, 
59~Cyg, FY~CMa \citep{pet08}, and $\phi$~Per \citep{gie98}, 
and there are some striking similarities among these binaries. 
Table 5 lists the orbital period and eccentricity, mass, 
effective temperature, and luminosity for each component of these systems. 
All have a low mass ratio, $M_2 / M_1 \approx 0.11$, and 
the temperatures of the respective components are very similar. 
Furthermore, the hot sdO component is a slow rotator in each case, 
suggesting that the progenitor had realized internal synchronous 
rotation with the orbit prior to the completion of mass transfer. 
The hot subdwarfs in these binaries are more massive and brighter
than the better known sdO stars in the field that represent a 
mixture of highly evolved, low mass stars \citep{heb09}. 
On the other hand, the subdwarfs in Be binaries have temperatures
and luminosities that are similar to those predicted by 
mass transfer models of binary evolution \citep{wel01}, 
and their presence is strong evidence that these Be stars 
were spun up by an earlier stage of mass transfer when 
the stars were much closer.  For example, if we assume 
conservative mass exchange, an initial mass ratio of $M_2 / M_1 = 1.25$
(typical of the systems modeled by \citealt{wel01}), and 
a final mass ratio of $M_2 / M_1 = 0.11$, then the current orbital  
periods should be $\approx 21 \times$ larger than their original periods. 
Thus, these systems probably began their lives as close pairs of B-stars. 

\placetable{tab5}      % Table 5 - Three Be + sdO systems 

We also list in Table~5 the orbital eccentricities of the Be + sdO 
binaries.  Only 59~Cyg has an eccentricity that is significantly 
different from zero, which is unusual because the intense tidal 
forces that the system experienced during large scale mass transfer
would tend to circularize the orbit.  We suspect that the massive 
third star in the system, Ab, may be a source of gravitational 
perturbations that have led to a secular increase of the eccentricity
following the conclusion of mass transfer \citep{for00}. 

The circumstellar disks of Be stars in Be + sdO systems are probably
truncated by the tidal influence of the companion.  We showed in Section~4
how the radius of the \ion{He}{1} $\lambda 6678$ emission region can be
estimated based upon Keplerian motion (see eq.~1).  The heated region 
probably forms near the outer disk boundary in the direction of the companion, 
and it is interesting to compare this emission radius with the Roche radius. 
We list in the final column of Table~5 the ratio of the emission radius 
$R_{\rm em}$ \citep{gie93,ste00,pet08} to the Roche radius at periastron
$(1-e) R_{\rm Roche}$ (derived from the mass ratio and the formula of 
\citealt{egg83}).  We see that in all three cases the disk extends to 
at least $78 - 92\%$ of the available Roche radius, so that in these
active Be stars the disks are probably as large as permitted by tidal 
limitations. 

The example of 59~Cyg demonstrates that hot subdwarf companions of Be stars
can be very difficult to detect.  Indeed, it is thanks to the tenacity 
of the {\it IUE} observers that a sufficiently large number of UV spectra 
were available to make detection possible through the use of 
cross-correlation and Doppler tomography techniques.  Thus, 
the question of how many Be stars may host hot subdwarf companions 
(and thus how many were spun up by mass transfer) remains 
very uncertain because of the observational difficulties surrounding 
detection.  Searching for such hot companions is of key importance to 
understand how Be stars became rapid rotators and to determine the 
numbers and kinds of their evolutionary progeny. 

%%%%%%%%%%%%%%%%%%%%%%%%%%%%%%%%%%%%%%%%%%%%%%%%%%%%%%%%%%%%%%% 
 
\acknowledgments 
 
The {\it IUE} spectra presented in this paper were obtained 
from the Mikulski Archive for Space Telescopes 
(MAST) at STScI.  STScI is operated by the Association of  
Universities for Research in Astronomy, Inc., under NASA  
contract NAS5-26555.  Support for MAST for non-HST data is  
provided by the NASA Office of Space Science via grant  
NNX09AF08G and by other grants and contracts. 
Many researchers contributed to the {\it IUE} spectroscopy 
used in this study, including programs by G.\ J.\ Peters, T.\ P.\ Snow, 
V.\ Doazan, J.\ M.\ Marlborough, J.\ M.\ Shull, P.\ L.\ Massey, 
A.\ J.\ Willis, C.\ A.\ Grady, C.-C.\ Wu, and H.\ F.\ Henrichs.
This work has made use of the BeSS database, operated at LESIA, 
Observatoire de Meudon, France: http://basebe.obspm.fr, and 
we are particularly grateful to Coralie Neiner and Benjamin Mauclaire
for the spectra of 59~Cyg that are collected at the BeSS. 
Our work was supported in part by NASA grant NNX10AD60G (GJP) and 
by the National Science Foundation under grant AST-1009080 (DRG). 
Institutional support has been provided from the GSU College 
of Arts and Sciences and the Research Program Enhancement 
fund of the Board of Regents of the University System of Georgia, 
administered through the GSU Office of the Vice President 
for Research, and by the USC Women in Science and Engineering 
(WiSE) program (GJP).  

{\it Facilities:} \facility{IUE, KPNO:CFT} 
 
%%%%%%%%%%%%%%%%%%%%%%%%%%%%%%%%%%%%%%%%%%%%%%%%%%%%%%%%%%%%%%% 
 
% References 

\clearpage

%%%%%%%%%%%%%%%%%%%%%%%%%%%%%%%%%%%%%%%%%%%%%%%%%%%%%%%%%%%%%%% 

% Tables  
 
% Table 1 
\begin{deluxetable}{lccccccccc}
\rotate 
\tabletypesize{\scriptsize} 
\tablewidth{0pc} 
\tablenum{1} 
\tablecaption{{\it IUE} Radial Velocity Measurements\label{tab1}} 
\tablehead{ 
\colhead{Date}             & 
\colhead{UT}               & 
\colhead{SWP}              & 
\colhead{Orbital}          & 
\colhead{$V_1$}            & 
\colhead{$\sigma_1$}       & 
\colhead{$(O-C)_1$}        & 
\colhead{$V_2$}            & 
\colhead{$\sigma_2$}       & 
\colhead{$(O-C)_2$}        \\  
\colhead{(HJD--2,400,000)} & 
\colhead{(yyyy-mm-dd)}     & 
\colhead{Number}           & 
\colhead{Phase}            & 
\colhead{(km s$^{-1}$)}    & 
\colhead{(km s$^{-1}$)}    & 
\colhead{(km s$^{-1}$)}    & 
\colhead{(km s$^{-1}$)}    & 
\colhead{(km s$^{-1}$)}    & 
\colhead{(km s$^{-1}$)}    \\
\colhead{(1)}              & 
\colhead{(2)}              & 
\colhead{(3)}              & 
\colhead{(4)}              & 
\colhead{(5)}              & 
\colhead{(6)}              & 
\colhead{(7)}              & 
\colhead{(8)}              & 
\colhead{(9)}              & 
\colhead{(10)}             
} 
\startdata 
43708.914 \dotfill & 1978-07-19 & \phn  2050 &  0.156 & 
\phn\phn     $  -3.4$ &  4.3 &          $ -15.1$ &
\nodata               & \nodata & \nodata       \\
43864.202 \dotfill & 1978-12-21 & \phn  3664 &  0.665 & 
\phn\phs     $  12.9$ &  3.2 & \phs     $  17.7$ &
\phn\phs     $  59.9$ &  6.7 & \phn     $  -1.2$ \\
44034.484 \dotfill & 1979-06-09 & \phn  5464 &  0.706 & 
\phn         $ -11.3$ &  3.8 & \phn     $  -4.5$ &
\nodata               & \nodata & \nodata       \\
44171.905 \dotfill & 1979-10-25 & \phn  6990 &  0.582 & 
\phn         $ -13.3$ &  3.2 &          $ -13.1$ &
\nodata               & \nodata & \nodata       \\
44277.719 \dotfill & 1980-02-08 & \phn  7894 &  0.336 & 
\phn\phs     $  17.6$ &  3.7 & \phn\phs $   5.8$ &
             $-118.6$ &  5.6 & \phn     $  -7.5$ \\
\enddata 
\tablecomments{A machine readable version of the full Table 1 is available 
in the electronic edition of the {\it Astrophysical Journal.}  A portion 
is shown here for guidance regarding its form and content.}
\end{deluxetable}
  
\newpage 
 
% Table 2 
\begin{deluxetable}{lccc} 
%\tabletypesize{\scriptsize} 
\tablewidth{0pc} 
\tablenum{2} 
\tablecaption{Orbital Elements for 59 Cyg\label{tab2}} 
\tablehead{ 
\colhead{Element} & 
\colhead{Harmanec et al.\ (2002)}  &
\colhead{Maintz (2003)} & 
\colhead{{\it IUE} ccfs}} 
\startdata 
$P$~(days)                \dotfill & $28.1971 \pm 0.0038$ & $28.192 \pm 0.004$ & $28.1871 \pm 0.0011$ \\ 
$T$ (HJD--2,400,000)      \dotfill & \nodata              & $51035.9 \pm 0.2$  & $45677.6 \pm 0.3$    \\ 
$T_{SC}$ (HJD--2,400,000) \dotfill & $50006.8 \pm 0.6$    & $51019.7 \pm 0.2$  & $45693.0 \pm 0.5$    \\ 
$K_1$ (km s$^{-1}$)       \dotfill & $13.0 \pm  1.0$      & $24.8 \pm  3.9$    & $11.7 \pm 0.9$       \\ 
$K_2$ (km s$^{-1}$)       \dotfill & \nodata              & $120.1 \pm 0.8$    & $121.3 \pm 1.1$      \\ 
$\gamma_1$ (km s$^{-1}$)  \dotfill & $-16.5 \pm  0.8$     & $+4.3 \pm 1.5$     & $+2.1 \pm 0.7$\tablenotemark{a}\\ 
$\gamma_2$ (km s$^{-1}$)  \dotfill & \nodata              & $-0.9 \pm 0.5$     & $-10.4 \pm 0.8$      \\ 
$e$                       \dotfill & 0\tablenotemark{b}   & $0.113 \pm 0.005$  & $0.141 \pm 0.008$    \\
$\omega_1$ (deg)          \dotfill & \nodata              & $293 \pm 3$        & $257 \pm 4$          \\
$M_2/M_1$                 \dotfill & \nodata              & $0.206$            & $0.097 \pm 0.008$    \\ 
$M_1\sin ^3 i$ ($M_\odot$)\dotfill & \nodata              & $7.23$             & $6.08 \pm 0.14$      \\ 
$M_2\sin ^3 i$ ($M_\odot$)\dotfill & \nodata              & $1.49$             & $0.59 \pm 0.05$      \\ 
$a\sin i$ ($R_\odot$)     \dotfill & \nodata              & $80.22$            & $73.3 \pm 0.8$       \\ 
rms$_1$ (km s$^{-1}$)     \dotfill & 3.1                  & \nodata            & 8.1                  \\ 
rms$_2$ (km s$^{-1}$)     \dotfill & \nodata              & \nodata            & 9.1                  \\ 
\enddata 
\tablenotetext{a}{Relative to the mean spectrum template.}
\tablenotetext{b}{Fixed.}
\end{deluxetable} 

\newpage 
 
% Table 3 
\begin{deluxetable}{lcc} 
%\tabletypesize{\scriptsize} 
\tablewidth{0pc} 
\tablenum{3} 
\tablecaption{Stellar Parameters for 59 Cyg\label{tab3}} 
\tablehead{ 
\colhead{Parameter} & 
\colhead{Primary}   & 
\colhead{Secondary} } 
\startdata 
$T_{\rm eff}$ (kK)      \dotfill  & $ 21.8 \pm 0.7 $  & $ 52.1 \pm 4.8 $ \\ 
$\log g$ (cgs)          \dotfill  & $ 3.78 \pm 0.09 $ & $ 5.0  \pm 1.0 $ \\ 
$V\sin i$ (km s$^{-1}$) \dotfill  & $ 379  \pm 27 $   & $ < 40 $         \\ 
$M/M_\odot$	        \dotfill  & $ 6.3 - 9.4 $     & $ 0.62 - 0.91  $ \\ 
$R/R_\odot$	        \dotfill  & $ 5.5 - 7.0 $     & $ 0.34 - 0.43  $ \\ 
\enddata 
\end{deluxetable} 

\newpage 
 
% Table 4 
\begin{deluxetable}{lcccc} 
%\tabletypesize{\scriptsize} 
\tablewidth{0pc} 
\tablenum{4} 
\tablecaption{\ion{He}{1} $\lambda 6678$ Spectral Observations\label{tab4}} 
\tablehead{ 
\colhead{Source} & 
\colhead{Dates}   & 
\colhead{$R=\lambda /\triangle\lambda$}   & 
\colhead{Number}   & 
\colhead{Observer} } 
\startdata 
KPNO/CF          & 1987 -- 1999   & 25800   & 22   & Peters  \\
BeSS/OHP/Elodie  & 2003           & 45000   &\phn8 & Neiner  \\
KPNO/CF          & 2004 -- 2008   &\phn9500 &\phn6 & Grundstrom  \\
BeSS/Castanet    & 2008 -- 2012   & 10000   &\phn4 & Buil  \\
BeSS/Revel       & 2009           & 10000   &\phn1 & Thizy   \\
BeSS/OVA         & 2010 -- 2012   & 10000   & 18   & Mauclaire  \\
\enddata 
\end{deluxetable} 

\newpage 
 
% Table 5 
\begin{deluxetable}{lccccccccc} 
\tabletypesize{\scriptsize} 
\tablewidth{0pc} 
\tablenum{5} 
\tablecaption{Parameters for Be + sdO Binaries\label{tab5}} 
\tablehead{ 
\colhead{Star} & 
\colhead{$P$}   & 
\colhead{}   & 
\colhead{$\log M/M_\odot$}   & 
\colhead{$\log M/M_\odot$}   & 
\colhead{$\log T_{\rm eff}$}   & 
\colhead{$\log T_{\rm eff}$}   & 
\colhead{$\log L/L_\odot$}  &
\colhead{$\log L/L_\odot$} &
\colhead{} \\
\colhead{Name} & 
\colhead{(d)}   & 
\colhead{$e$}   & 
\colhead{(Be)}   & 
\colhead{(sdO)}   & 
\colhead{(Be)}   & 
\colhead{(sdO)}   & 
\colhead{(Be)}  &
\colhead{(sdO)} &
\colhead{${R_{\rm em}} / [(1-e) R_{\rm Roche}]$}
} 
\startdata 
59 Cyg     \dotfill & \phn 28.2  & 0.14 &\phs 0.9 &    $-0.1$ & 4.3 & 4.7 & 3.9 & 3.0 & 0.88 \\    
FY CMa     \dotfill & \phn 37.3  & 0.00 &\phs 1.1 & \phs 0.1  & 4.4 & 4.7 & 4.2 & 3.3 & 0.92 \\
$\phi$ Per \dotfill &     126.7  & 0.00 &\phs 1.0 & \phs 0.1  & 4.4 & 4.7 & 4.3 & 4.1 & 0.78 \\
\enddata 
\end{deluxetable} 

\clearpage

%%%%%%%%%%%%%%%%%%%%%%%%%%%%%%%%%%%%%%%%%%%%%%%%%%%%%%%%%%%%%%% 

% Figures 

% Figure 1 
 
% Figure 2 
 
% Figure 3 
 
% Figure 4 
 
% Figure 5 
 
% Figure 6 
 
% Figure 7 

% Figure 8 
\begin{figure} 
\begin{center} 
{\includegraphics[height=16cm]{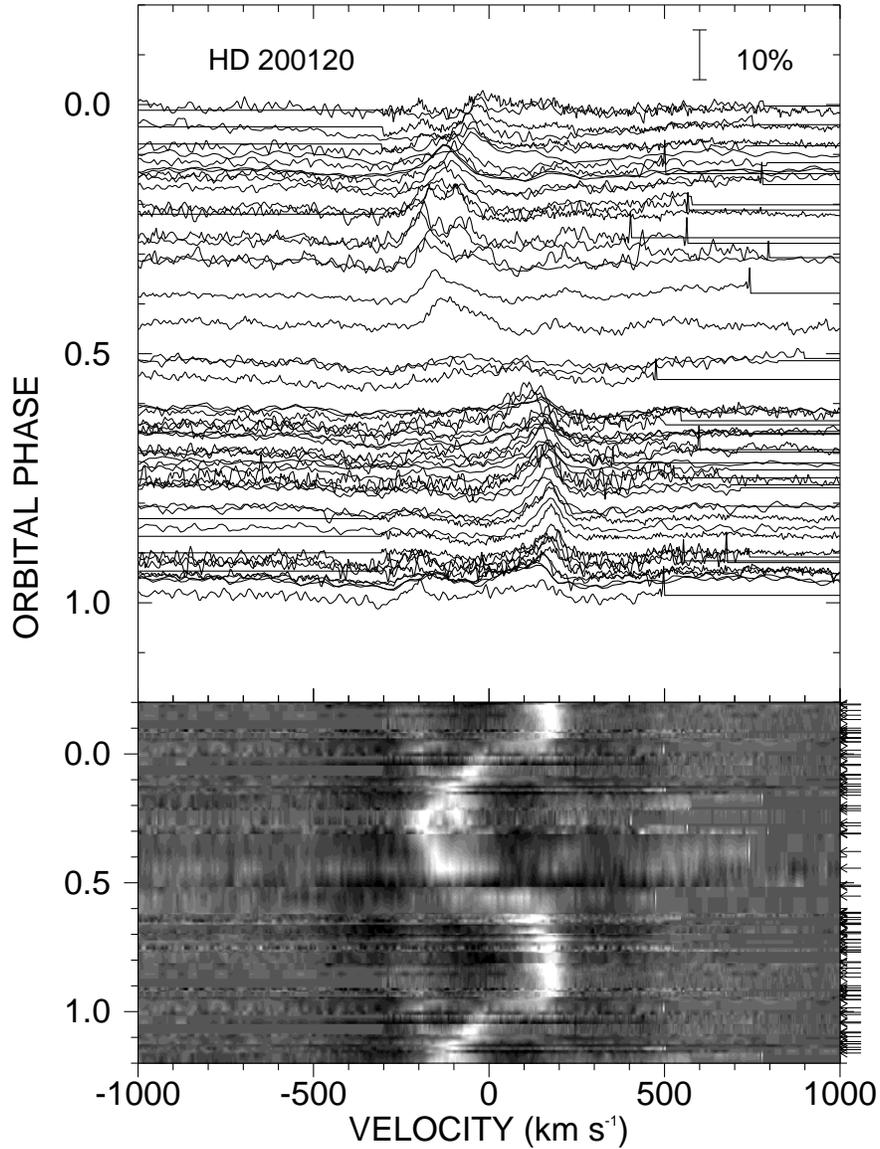}}
\end{center} 
\caption{The orbital phase variations in the \ion{He}{1} $\lambda 6678$
emission line in the spectra of 59~Cyg are shown in  
linear plots ({\it top panel}) and as a gray-scale image ({\it lower panel}).  
The intensity in the gray-scale image is assigned according to
the flux between the minimum (dark) and maximum (bright) observed values.  
The intensity between observed spectra is calculated by a linear interpolation  
between the closest observed phases (shown by arrows along the right axis). 
The scale bar at top right indicates the spectral flux 
relative to the local continuum flux.
\label{fig8}} 
\end{figure}

\end{document}